\documentclass[twocolumn, prb]{revtex4}

\usepackage{graphicx}
\usepackage{dcolumn}
\usepackage{bm}

\begin{document}

\title{Thermoelectric enhancement in PbTe with K, Na co-doping from tuning the
interaction of the light and heavy hole valence bands}

\author{John Androulakis}
\affiliation{Department of Chemistry, Northwestern University, Evanston, Illinois, 60208, USA}

\author{Iliya Todorov}
\author{Duck-Young Chung}
\affiliation{Materials Science Division, Argonne National Laboratory, Argonne, Illinois, 60439, USA}

\author{Sedat Ballikaya}
\affiliation{Department of Physics, University of Michigan, Ann Arbor, Michigan 48109, USA \\
Department of Physics, University of Istanbul, 34134 Vezneciler, Istanbul, Turkey}

\author{Guoyu Wang}
\author{Ctirad Uher}
\affiliation {Department of Physics, University of Michigan, Ann Arbor, Michigan 48109, USA}

\author{Mercouri Kanatzidis}
\email{m-kanatzidis@northwestern.edu}
\affiliation{Department of Chemistry, Northwestern University, Evanston, Illinois, 60208, USA\\
Materials Science Division, Argonne National Laboratory, Argonne, Illinois, 60439, USA}

\date{\today}

\begin{abstract}
The effect of K and K-Na substitution for Pb atoms in the rock salt
lattice of PbTe was investigated to test a hypothesis for development
of resonant states in the valence band that may enhance the
thermoelectric power. We combined high temperature Hall-effect,
electrical conductivity and thermal conductivity measurements to show
that K-Na co-doping do not form resonance states but2 can control
the energy difference of the maxima of the two primary valence
sub-bands in PbTe. This leads to an enhanced interband interaction with
rising temperature and a significant rise in the thermoelectric figure
of merit of p-type PbTe. The experimental data can be explained by a
combination of a single and two-band model for the valence band of PbTe
depending on hole density that varies in the range of $1-15 \times 10^{19} cm^{-3}$.
\end{abstract}

\maketitle

\section{\label{sec:intr} Introduction}
The primary ellipsoidal valence and conduction bands in the ground state of PbTe consist of four valleys contributing to a substantially high Seebeck coefficient which makes it an outstanding thermoelectric material for fundamental investigations and for power generation in the temperature regime 600-800 K.\cite{Dughaish} Recently, compelling improvements of the thermoelectric figure of merit have been demonstrated through strong reduction of the lattice thermal conductivity achieved from nano-structuring of the PbTe matrix.\cite{Sootsman, Kanatzidis} The thermodynamically driven formation of nanostructure rich in Ag-Sb, Na-Sb, Ag-Sn-Sb, and S that are embedded endotaxially in the PbTe matrix have shown reduced lattice thermal conductivity while at the same time not substantially degrading the beneficial electrical properties of pure PbTe.\cite{Hsu, Poudeu, AndroulakisHsu, Androulakis, Hoang}  The figure of merit is expressed through the dimensionless quantity $ZT = \frac{S^{2} \sigma T}{\kappa}$, S being the Seebeck coefficient, $\sigma$ the electrical conductivity and $\kappa$ the total thermal conductivity comprising contributions of the lattice phonons and charge carriers. The transport parameters entering the figure of merit strongly depend on temperature and different materials are considered for different temperature regimes.\cite{Kanatzidis, Snyder} 

The maximum thermoelectric figure of merit, $Z_{max}$, that can be attained at a given temperature T, depends on a quite diverse set of materials parameters according to Eq. \ref{eq:1}:\cite{Sootsman}

\begin{eqnarray}
 Z_{\mathit{max}}=\frac{\mathit{\gamma e}^{(r+1/2)}T^{3/2}\tau
_{z}\sqrt{\frac{m_{x}m_{y}}{m_{z}}}}{\kappa _{\mathit{lat}}}
\label{eq:1}
\end{eqnarray}

In the above, $\kappa_{lat}$ is the thermal conductivity of the lattice, $m_{i}$ is the effective mass in the i-th direction in the crystal, $\tau_{z}$ is the relaxation time of the carriers moving in the direction of the current flow (assumed to be $z$) and $\gamma$ is the degeneracy of the band extrema near the Fermi energy. It follows that for an isotropic system, high $\gamma$ and low $\kappa_{lat}$ values contribute significantly to high figure of merit.   

From a practical point of view the lattice thermal conductivity, $\kappa_{lat}$, of PbTe via nanostructuring and alloying is reaching its low limit. Therefore, further improvements in the ZT of PbTe can be sought through large enhancements of the efficiency of charge transport quantified by the product S$^{2}\sigma$, also known as the power factor. One approach to enhancing the power factor in PbTe has been described by Martin et al \cite{Martin} who have studied p-type PbTe nanocomposites and found that grain boundary scattering dominates charge transport acting as an energy filtering effect that enhances thermopower. Furthermore, Faleev and L\'eonard \cite{Faleev} have suggested enhancements in the power factor in the presence of metallic nanoinclusions. Again, an energy filtering effect is invoked due to band bending at the semiconductor-metal interface that considerably modifies the electronic scattering time.  Recently, Heremans et al \cite{Heremans} have investigated distortions in the density of states (DOS) of Tl-doped PbTe that can enhance the Seebeck coefficient, and hence the power factor, leading to increased figure of merit.\cite{Kaidanov} Such distortions in the DOS are known as resonant states that induce Fermi level pinning and have been observed in a number of narrow band gap chalcogenide semiconductors such as GeTe and SnTe.\cite{Bushmarina, Berezin, Story} The issue of the distortions in DOS and the presence of resonance states is still an open question and it has been recently discussed in detail in the theoretical work of Singh.\cite{Singh}

In general, dopants in narrow band gap semiconductors do not form mid-gap states.\cite{Pantelides} Rather, impurity states outside the energy gap, also known as deep defect states, are formed and interact strongly with the lattice in a way that considerable distortions, not always resonant-like, in the shape and energy distribution of the DOS can develop.\cite{Ahmad} In order for resonant states to benefit the thermoelectric properties they should have a sharp distribution in energy, $\varepsilon$, and preferably be located in energy span within 0.5eV from the Fermi level in order to be accessible by usual thermoelectric transport processes.\cite{Nemov-Ravich} Since there is no experimental way to predict {\it a priori} the formation and behavior of resonating states in PbTe, it is evident that the selection of appropriate chemical substitutions for either Pb or Te could benefit from theoretical calculations. 

Following the theoretical work of Ahmad et al \cite{Ahmad} which predicts a resonant-like DOS distortion in p-type Pb$_{1-x}$A$_{x}$Te, (where A=K, Rb, Cs, but not Na) we investigated doped PbTe giving particular emphasis on K, since our data suggest that the larger Rb and Cs ions appear to not substitute effectively for Pb. Furthermore, following the relevant theory on resonance scattering we doped PbTe concurrently with K and Na, the former intended as a possible resonance state and the latter, being an electrically active dopant that yields no impurity states, to tune the Fermi energy and to check for Fermi level pinning.\cite{Kaidanov, Nemov-Ravich} We have also studied control samples without K and containing Na only as a dopant. Although, our experimental results cannot support the presence of resonant scattering in our samples, we did observe enhancements in the power factor of the specimens at high temperature. We discuss our findings with respect to the high temperature electronic band structure of PbTe that takes into account the coexistence of two valence bands, a light- and a heavy-hole band, and the interaction between them which can be tuned by the presence of K in the lattice. The interplay between the two types of valence bands can be tuned with K-Na codoping to create a sharply changing DOS near the Fermi energy in a fashion similar to a resonant state. In this paper we combine, X-ray diffraction with high temperature Hall-effect, electrical conductivity, thermopower, and thermal conductivity measurements to show that codoping of PbTe with K and Na is an efficient way to control the energy separation between the maxima of the valence bands in order to achieve an effective Fermi level pinning, at a given temperature, with respect to the carrier concentration. We show that, in effect, this creates a resonant-like modification in the DOS which can be tuned by the K ions although these ions by themselves do not form a resonant level. A maximum $ZT \sim$1.3 at 700 K was observed as a result of increased power factor rather than reduced thermal conductivity.

\section{\label{sec:expt} Experimental Details}

Samples with the chemical formulas, Pb$_{1-x}$Na$_{x}$Te, Pb$_{1-x}$K$_{x}$Te and Pb$_{0.9875-x}$K$_{0.0125}$Na$_{x}$Te were produced by reacting high purity metals (American Elements Pb 99.99\% , 5N Plus Te 99.999\%, Aldrich K $>$99.95\%, Aldrich Na $>$99.95\%) as starting materials. (For simplicity, in the following we adopt the notation:  Pb$_{1-x}$Na$_{x}$Te $\rightarrow$ PbTe:Na x\%, Pb$_{1-x}$K$_{x}$Te $\rightarrow$ PbTe:K x\%, Pb$_{0.9875-x}$K$_{0.0125}$Na$_{x}$Te $\rightarrow$ PbTe:K 1.25\%:Na x\%.). Approximately 15mg of K and 4.5-12 mg of Na (per 10g of product) were first loaded in carbon coated silica tubes inside a glove box under nitrogen atmosphere. The K and Na mass was used as a reference to calculate appropriate quantities of Pb ($\sim$6.5g per 10g of product) and Te ($\sim$3.8g per 10g of product) that were added in the respective tube. All tubes were subsequently evacuated and fused. The tubes were fired at 1050 $^{o}$C over 12 hours and soaked at that temperature for 4 h before rapidly cooling them to room temperature over a period of 3 h. All sample ingots produced are metallic silvery gray in color and are strong enough to undergo cutting and polishing. The bottom part of the ingot was used in $\kappa$ measurements while a piece right above that was bar-shaped for S and $\sigma$ measurements in a way that the longer direction coincides with the direction in which $\kappa$ was measured. 

Powder x-ray diffraction studies, performed with an INEL diffractometer equipped with Cu Ka radiation and a position sensitive detector, showed no indication for a second phase within the resolution of the instrument. The diffractometer was calibrated for each sample against high purity Si. The observed diffractograms were indexed in the $Fm$\={3}$m$ space group and the 2$\theta$ position of the (200) peak was used to calculate the lattice parameters of the samples.
	
A NETZSCH LFA 457 Microflash instrument was used to determine the thermal diffusivity of coin-shaped samples with a typical diameter of 7.9 mm and thickkess of $\sim$ 2mm. The thermal conductivity, $\kappa$, was then calculated by the relation $\kappa=DC_{p}\rho$, where D is the thermal diffusivity, $C_{p}$ is the heat capacity under constant pressure and $\rho$ is the mass density of the specimens. To increase the accuracy of thermal conductivity data the density of the samples was extracted by measuring their volume in an Accupyc-1340 pycnometer. This method yields a mass density 8.1-8.21 g/cm$^{3}$, that practically coincides with the theoretical one, and yields 5-10\% better accuracy compared to the density calculated using the geometrical dimensions of the specimens. Furthermore, Cp was carefully adjusted to match $\sim$ 98\% that of PbTe according to the data by Blachnik et al. \cite{Blachnik}. 

The electrical conductivity, $\sigma$, and Seebeck coefficient, S, were measured simultaneously under a helium atmosphere ($\sim$ 0.1 atm) from room temperature to $\sim$ 700 K using a ULVAC-RIKO ZEM-3 system.  The Hall coefficient was measured in a home-made high temperature apparatus, which provides a working range from 300 K to 873 K.  The samples were press-mounted and protected with argon gas to avoid possible oxidization at high temperature.  The Hall resistance was monitored with a Linear Research AC Resistance Bridge (LR-700), and the data were taken in a field of $\pm$ 1 T provided by an air-bore Oxford Superconducting Magnet. The hole densities, p, at room temperature were extracted on the assumption of a single band which gives for the Hall coefficient, $R=1/pe$, where $e$ is the electron charge.

\section{\label{sec:res-disc} Results and Discussion}

Resonance scattering is a form of selective scattering where carriers exhibiting relaxation times and mean free paths within certain limits are scattered preferentially compared to carriers whose energies lie outside these limits.\cite{Ravich-CRC} This selectivity, in effect, has a strong impact on the electronic properties which can lead to an enhancement of S as a function of carrier concentration at a given temperature when compared to the same property in the absence of resonating scattering. Usually, the reference temperature is 300K where S of PbTe follows the so-called Pisarenko formula (non-degenerate):
\begin{eqnarray}
 S=\frac{k_{B}}{e}[\frac{5}{2}+r+\ln
(\frac{2}{p}(\frac{m_{d}^{\text{*}}k_{B}T}{2\pi \hbar })^{3/2})]
\label{eq:2}
\end{eqnarray}
where $r$ is the scattering parameter that takes the value -1/2 for acoustic phonon scattering, $m_{d}^{\text{*}}$ is the DOS effective mass, $p$ the hole density, $k_{B}$ the Boltzmann constant, $\hbar$ the reduced Planck's constant, e the electronic charge and T the temperature. We note that the high symmetry of the valence and conduction bands in PbTe makes Eq. (2) equivalent for both n- and p-type samples with the same {\it non-degenerate} carrier densities where electron- acoustic phonon scattering is dominant. \cite{Jaworski} A distortion in the DOS would result in a deviation from the Pisarenko line expressed as an enhancement in the S values at T=300K for a particular hole density.\cite{Heremans}

Considerable deviations from Eq. (2) can also occur in highly degenerate p-type PbTe as a result of the special nature of the valence band structure of PbTe. Allgaier\cite{Allgaier} has shown that Hall data in this region can be explained by assuming a more complex valence band structure. In this model a principal highly non-parabolic band, hitherto called band 1 (light hole), with energy extrema located at the $\langle 111 \rangle$ directions in k-space coexists with a parabolic band, hitherto called band 2 (heavy hole). These bands have substantially different properties; band 1 is characterized by high energy dispersion, a light effective mass ($ \sim 0.2 m_{0}$), and a lower density of electronic states (DOS) compared to band 2 that exhibits a heavy effective mass (1.2-2$m_{0}$) has a high DOS and weaker dispersion.\cite{AVDKR} At T=0 K, the top of band 2 is located 0.14-0.17 eV below that of band 1.\cite{AVDKR} Therefore, increasing carrier doping moves the Fermi level deeper in energy and closer to band 2. 
 
\begin{figure}[hb]
\centerline{
\includegraphics[width=0.40\textwidth]{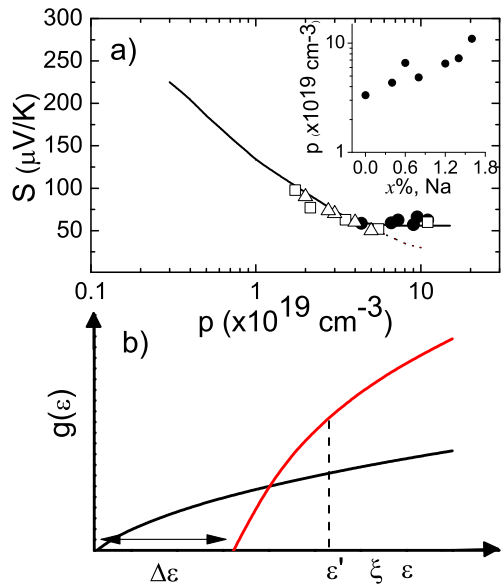}
}
\vspace{0.5cm}
\caption{\label{fig:Figure 1} a) Seebeck coefficient as a function of carrier concentration (Pisarenko plot) at T=300K for PbTe samples doped with Na (open squares), K (open triangles) and co-doped with K-Na (filled circles). The co-doped samples correspond to 1.25\% K and varying Na concentration. The black solid line is the theoretical Pisarenko line that takes into account the valence band structure of PbTe. Notice the change in curvature around $2 \times 10^{19} cm^{-3}$. The dashed line corresponds to a single non-parabolic band line with the assumption of electron-phonon scattering and holds well for non-degenerate doping. The inset to Fig. 1a shows the hole density as a function of increasing Na concentration for co-doped samples of PbTe:K 1.24\%:Na $x$\%. b)Schematic of the DOS, $g(\varepsilon)$, versus energy, $\varepsilon$, that shows the band overlap at an arbitrary band maxima energy separation, $\Delta\varepsilon$. When the chemical potential, $\xi$, is sufficiently high there is a sudden increase in the DOS which acts to effectively pin the Fermi energy with any further increase of the hole density.
}
\end{figure}

The top of band 2 moves up in energy with rising temperature and at $\sim$ 450-500 K can reach the top of band 1. At room temperature band 2 can be accessed by a high level of acceptor doping. When both bands are populated with carriers the thermopower reflects the contribution of both bands according to the formula:
\begin{eqnarray}
 S=\frac{S_{1}b_{p}p_{1}+S_{2}p_{2}}{b_{p}p_{1}+p_{2}}
\label{eq:3}
\end{eqnarray}
where S$_{1}$, S$_{2}$ denote individual thermopower contributions from the bands 1 and 2, p$_{1}$ and p$_{2}$  are the hole densities of bands 1 and 2, and b$_{p}$ is defined as the mobility ratio of band 2 with respect to band 1. The results of such analysis at 300 K yield a constant S of $\sim$ 56 $\mu$V/K for hole densities $> 3 \times 10^{19} cm^{-3}$.\cite{AVDKR}

Figure 1a displays the Pisarenko plot, S measured as a function of hole density, for PbTe:Na $x$\% (0.1$\leq x\leq$ 2.5) specimens (squares), PbTe:K $x$\% (0$\leq x\leq$ 2.5) specimens (triangles) and PbTe:K 1.25\%:Na $x$\% (0.4$\leq x\leq$2.5) specimens (filled circles) at T=300 K. The dotted black line is calculated from Eq. 2 and the solid line is the calculated line at the degenerate limit.\cite{AVDKR} Evidently, at room temperature there is no enhancement in S in contrast to the case of Tl doped PbTe and therefore resonance scattering most probably does not take place in the K doped samples. The inset of Fig. 1 shows the monotonic increase of hole density with increasing Na doping in PbTe:K 1.25\%:Na $x$\%. Therefore, there is an apparent independence of S with increasing carrier concentration that is consistent with Fermi level pinning and can be qualitatively explained by the schematic shown in Fig 1b. The non-parabolic DOS (black line), $g(\varepsilon)$, is overlapped at an energy $\varepsilon '$ by the states emanating from band 2 (red line). The narrower dispersion of band 2 results in a considerable enhancement of $g(\varepsilon)$ for $\varepsilon > \varepsilon '$ and therefore the reduced Fermi energy, $\xi$, is only weakly affected ('effective pinning') by changes in doping in this region at a given temperature.  Increasing temperature brings about large changes in the DOS since it reduces the energy difference, $\Delta\varepsilon$, of the two band extrema with respect to the Fermi energy. With rising temperature there is lattice thermal expansion which causes the heavy hole band 2 to rise in energy whereas the light hole band 1 falls in energy. Therefore, $\Delta\varepsilon$, can be controlled not only with temperature but also with engineered lattice expansion of the materials.  In the PbTe:K, and PbTe:K:Na samples we find that there is lattice expansion compared to pure PbTe and PbTe:Na, Fig. 2, consistent with the ionic radii of six-coordinate K$^{+}$ (133 pm), Na$^{+}$ (102 pm), and Pb$^{2+}$ (119 pm).\cite{Shannon}

\begin{figure}[hb]
\centerline{
\includegraphics[width=0.50\textwidth]{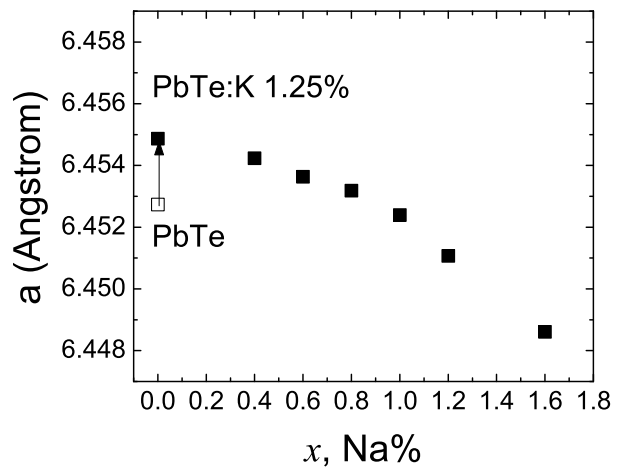}
}
\vspace{0.5cm}
\caption{\label{fig:Figure 2} Lattice parameters at room temperature of PbTe:K 1.25\%:Na $x$\% as a function of increasing Na content. The rapid decrease above 1\% Na can be attributed to the expulsion of K atoms from the structure by the smaller Na ions. Notice the expansion of the lattice with 1.25\% K doping with respect to pure PbTe (open square).
}
\end{figure}

Next we present results and discuss the effect of temperature on the band structure and the thermoelectric properties of the co-doped samples in comparison to the control samples, i.e. PbTe doped with Na only. 

Figure 3(a) depicts $\sigma$ as a function of temperature for selected specimens of the PbTe:K 1.25\%:Na $x$\% series of samples. The nominal compositions (exact values of $x$) are shown in the figure. All samples exhibit high electrical conductivity at room temperature that decreases with increasing temperature. 

To assess the scattering rate we assumed a single power law dependence of $\sigma \sim 1/T^{\delta}$. Subsequently all data sets were scaled in $1/T^{\delta}$ and fitted to a straight line. The extracted $\delta$ values as a function of Na concentration are given in Fig. 3(b). The values deviate from the normal 5/2 law\cite{Ioffe} with increasing $x$ up to 1\% and then the trend is reversed with a further increase in the Na content. This weaker power law dependence effectively maintains high electrical conductivity at high temperatures. It is worth pointing that $\sigma$ of the 0\% Na sample decreases faster compared to all other samples reaching a value of 168 S/cm at 720K with a $\delta$ exponent of $\sim$2.7. By comparison PbTe:K $x$\% samples also exhibit $\delta$ values as high as 3 depending on $x$. Samples co-doped with K and Na, however, exhibit $\sigma$ values around 700K that range from 260 to 360 S/cm depending on the Na content with considerably lower $\delta$ exponents of 2.6-1.3, see Fig. 3b. We note that a $\delta$ value as low as 1.4 has been observed by Kolomoets et al \cite{Kolomoets1} in two highly degenerate Na only doped PbTe samples ($8 \times 10^{19}$ and $1.4 \times 10^{20} cm^{-3}$ respectively) and this was explained in terms of interband scattering, i.e., the transfer of holes from band 1 to band 2. Since charge scattering in PbTe is dominated by phonons,\cite{Vineis} the reduction in the power law dependence of $\sigma$ in the presence of K (expanded lattice compared to Na doping) may suggest significant phonon softening.  

\begin{figure}[hb]
\centerline{
\includegraphics[width=0.40\textwidth]{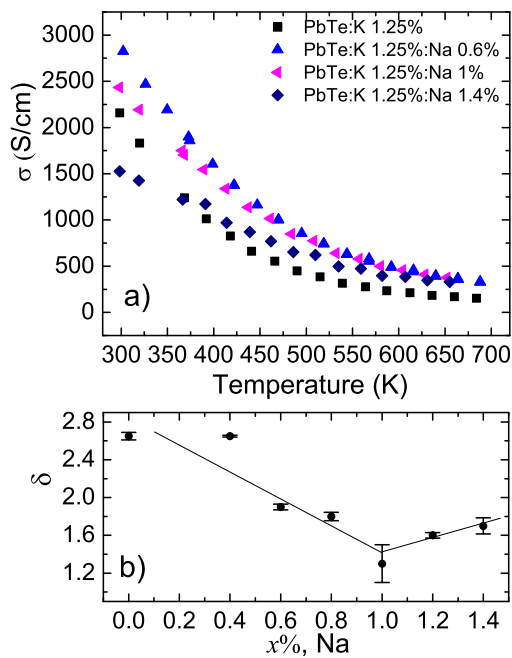}
}
\vspace{0.5cm}
\caption{\label{fig:Figure 3} a) Electrical conductivity, $\sigma$, as a function of temperature for selected samples co-doped with K and Na. The nominal compositions are given. It is noteworthy that $\sigma$ remains high at elevated temperatures. Notice that the curve for 0\% Na drops much faster.  b) Power law exponents, $\delta$, of the electrical conductivity as a function of Na content for all co-doped samples. A minimum is reached at $x \sim$1\%. The solid lines are guides to the eye.
}
\end{figure}

The Seebeck coefficient of the co-doped specimens PbTe:K 1.25\%:Na $x$\% as a function of temperature is shown in Fig 4. At 300K a similar magnitude $\sim$ 60$\mu$V/K for all $x$ values is observed as a result of degeneracy that pushes the Fermi energy within band 2. For T$>$300 K, S increases monotonically up to 700 K for all $x$ values and the S-T curves remain below that of the purely K doped specimen. The values measured at 700 K range from 240 to 300 $\mu$V/K depending on $x$. It is noteworthy that for $x>$0.8\%, the S curves coincide in temperature suggesting that the physical picture of Fig. 1(b) may also be in effect at high temperatures. In such a case the determining quantity for a high power factor is $\sigma$, which depends on factors affecting the carrier mobility e.g. defects, grain boundaries etc.
\begin{figure}[hb]
\centerline{
\includegraphics[width=0.50\textwidth]{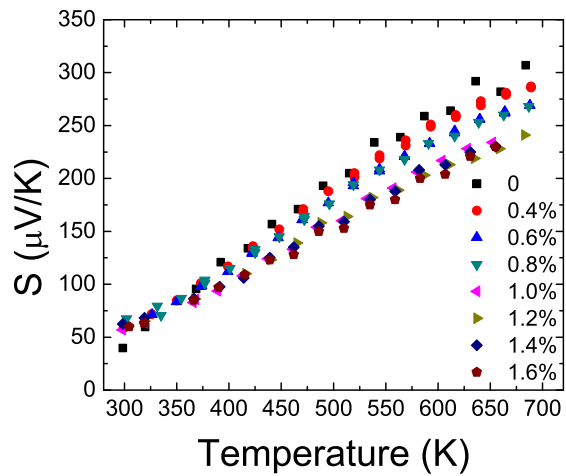}
}
\vspace{0.5cm}
\caption{\label{fig:Figure 4} Seebeck coefficient, S, as a function of temperature for PbTe:K 1.25\%:Na $x$\%  samples, with $x$ noted on the figure. Notice that at room temperature all S values for the co-doped samples coincide. At higher temperatures the curves scale with Na concentration (lower S for higher Na content) reflecting the corresponding increase in hole density up to $x \sim$ 0.8\%. For $x \geq$1\% all S-T curves coincide pointing to an “effective pinning” mechanism as explained in Fig 1b. 
}
\end{figure}

Figure 5 presents the Hall coefficient, $R_{H}$, as a function of T for selected specimens of the PbTe:K 1.25\%:Na $x$\% series. $R_{H}$ increases with increasing T up to a maximum value and then falls with a further increase in temperature. This maximum can be qualitatively explained on a two-band Hall coefficient model as a function of temperature.\cite{Allgaier-2} In such a model Allgaier assumed that the ratio of the Hall factors of the two bands is almost unity and that $b_{p} < < 1$ expressed the Hall coefficient as:
\begin{eqnarray}
\frac{R-R_{0}}{R_{0}}=\frac{(1-b_{p})^{2}t}{(1+b_{p}t)^{2}}
\label{eq:4}
\end{eqnarray}
Here, $t=p_{2}/p_{1}$  is related to the DOS and the $F_{1/2}(\xi)$ Fermi integral and $R_{0}=r_{1}/e(p_{2}+p_{1})$, i.e., the Hall coefficient at sufficiently low T where all holes are transferred to band 1. Furthermore $\Delta \varepsilon$ is related to the ratio of DOS of the two bands, $P$, and the Fermi energy at T=0 K through the formula:
\begin{eqnarray}
\frac{2}{3}\frac{E_{F}}{\mathit{\Delta \varepsilon }}\xi
^{3/2}=F_{1/2}(\xi )+\mathit{PF}_{1/2}(\xi -\frac{\mathit{\Delta
\varepsilon }}{k_{B}T})
\label{eq:5}
\end{eqnarray}
 
\begin{figure}[hb]
\centerline{
\includegraphics[width=0.50\textwidth]{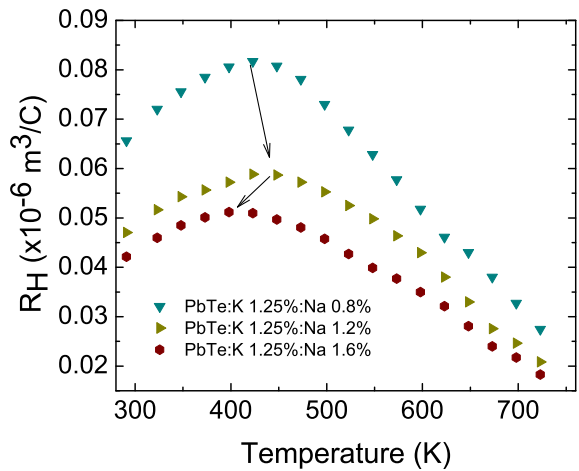}
}
\vspace{0.5cm}
\caption{\label{fig:Figure 5} Hall coefficient, $R_{H}$, as a function of temperature for three selected co-doped samples with nominal compositions indicated on the graph. The arrows show that the position of the peak shifts to higher temperatures for the sample with doping in the vicinity of 1.2\% Na, i.e. where the minimum power law exponent in electrical conductivity is observed. 
}
\end{figure}

Allgaier's analysis emphasizes the importance of $P$, $b_{p}$, and $\Delta \varepsilon$. It is clear that $\Delta \varepsilon$ does not determine whether a maximum occurs, however it does specify where it occurs. The maximum in $R_{H}(T)$ shifts from 415K for $x$=0.8\% to 430 K for $x$=1.2\% before moving back to 410K for $x$=1.6\%. We relate the shift to lower temperatures in the $R_{H}$ maximum for the $x$=1.6\% sample with its decreased lattice parameter compared to pure PbTe (see Fig. 2) that forces the lattice to restore PbTe:Na-like behavior. Such an effect would also explain the slight increase in $\delta$ observed for $x>$1\%.   

The changing position of the $R_{H}$ maximum is a clear indication of energy shift in the band positions and therefore in $\Delta \varepsilon$. Further within this framework and for degenerate carrier densities Allgaier argued that $R_{H}$ should scale as $T^{3/2}$ as the result of classical statistics dominating in band 2. The slope of the $R_{H}$ vs $T^{3/2}$ is related to $\Delta \varepsilon$ and an estimate of the temperature dependence can be extracted.  Indeed, our data above the peak scale linearly in $T^{3/2}$. Similar dependence was also observed for PbTe:Na samples. However the K-Na co-doped samples of similar carrier density to Na-only samples exhibit slopes that are an order of magnitude lower. Therefore, we conclude that co-1doping affects not only $\Delta \varepsilon$ but also its temperature dependence making band 2 rise in energy at a much slower rate with increasing temperature.    

We note that lattice expansion in PbTe causes band 1 to decrease and band 2 to increase in energy which effectively diminishes $\Delta \varepsilon$ between the valence bands 1 and 2. At T$>$400K as the lattice continues to expand it places band 2 higher in energy than band 1 and band 2 completely dominates transport.\cite{Kolomoets1} The K-Na co-doping where Na is varied while K is kept constant allows the precise control of the Fermi level and opens a wider temperature range where bands 1 and 2 are forced to interact. Since both the total S and total $\sigma$ in the two valence band picture are weighted quantities (see Eq. 3), the interaction of the two bands for a wider temperature range implies higher power factors. This is because S is almost constant and independent of doping while the interband scattering mechanism lowers the exponent $\delta$ in the temperature dependence of the electrical conductivity. 

Figures 6(a) and (b) compare power factors between co-doped samples and Na-only or K-only samples with similar carrier densities at room temperature. Figure 6(a) refers to samples exhibiting carrier densities in the range $4-5 \times 10^{19} cm^{-3}$ and Fig. 6(b) refers to samples with carrier densities of $9-10 \times 10^{20} cm^{-3}$. Increasing K concentration in PbTe does not affect the carrier density beyond $5-6 \times 10^{19} cm^{-3}$, an effect attributed to the solubility limit of K. Therefore Fig. 6(b) compares co-doped samples with Na-only doped samples.  Evidently, the co-doped samples exhibit enhanced power factors in both cases. We note that for the heavily doped samples (Fig. 6(b)) at 300K the PbTe:Na sample has the highest power factor. However the power factor of the co-doped specimen exhibits a steeper slope in temperature and at T$\sim$400K it has already exceeded that of PbTe:Na. For T$>$400K the power factor of the co-doped samples is consistently 15-25\% higher (depending on temperature). The maximum power factor achieved at 700K is $S^{2} \sigma_{(700K)} \sim 23.5 \mu W/cmK^{2}$. It should also be noted that the power factor of the PbTe:Na ($ \sim 10^{20} cm^{-3}$) sample benefits primarily from a high electrical conductivity and not from a high Seebeck coefficient which at 700 K is $\sim$210 $\mu V/K$.

\begin{figure}[hb]
\centerline{
\includegraphics[width=0.40\textwidth]{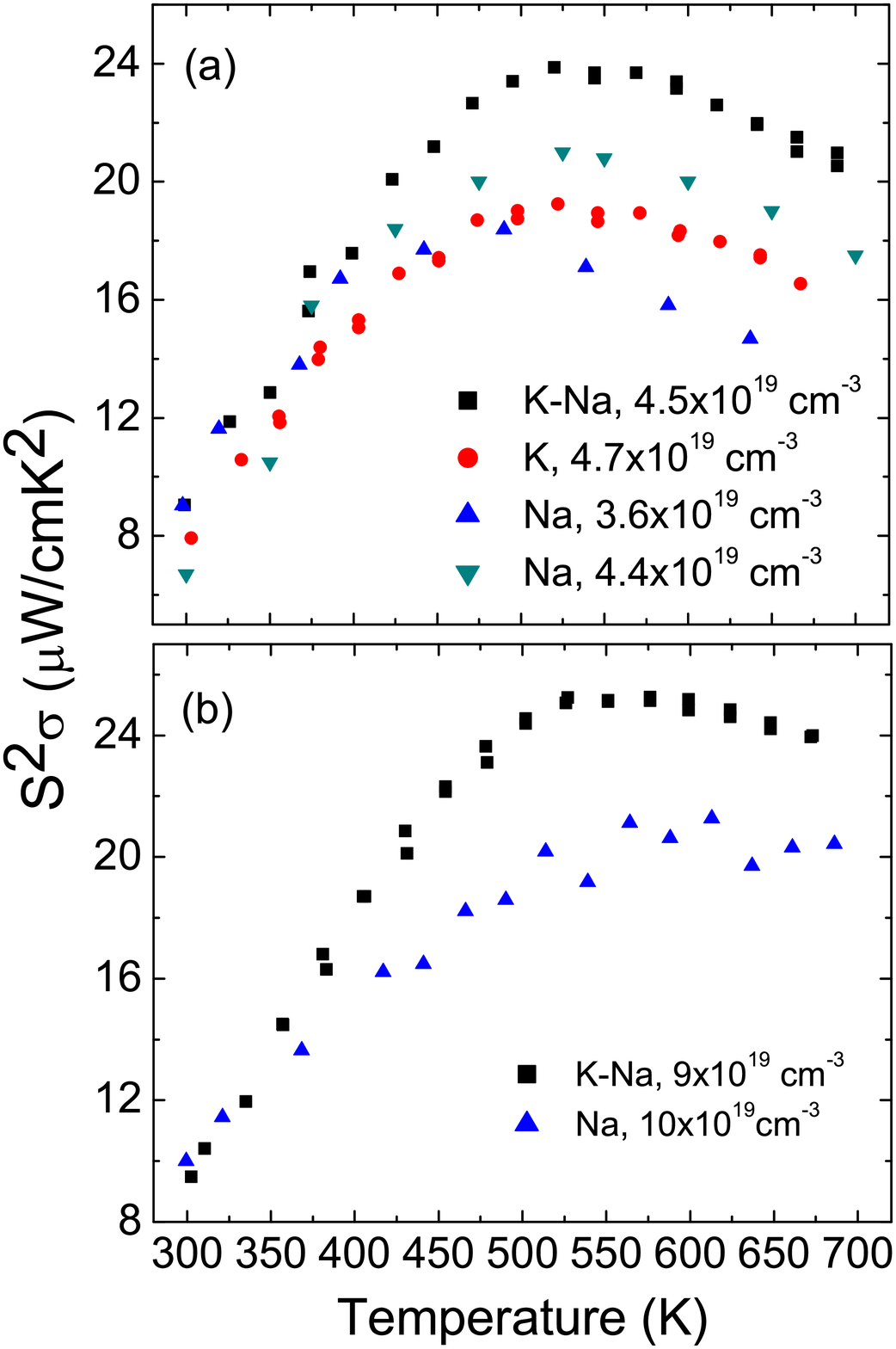}
}
\vspace{0.5cm}
\caption{\label{fig:Figure 6} (a) Power factors as a function of temperature for PbTe:K, PbTe:Na and PbTe:K:Na samples corresponding to a room temperature hole density of $ \sim 4.5 \times 10^{19} cm^{-3}$. The co-doped sample outperforms the PbTe:Na and PbTe:K samples for T$>$400K. (b) Power factors for PbTe:Na and PbTe:K:Na samples corresponding to a room temperature hole density of $ \sim 9-10 \times 10^{19} cm^{-3}$. Clearly, the co-doped sample exhibits a higher power factor for T$>$400 K reaching 24 $\mu$W/cmK$^{2}$ at 700K. 
}
\end{figure}

Figure 7 shows the measured total thermal conductivity, $\kappa_{tot}$, as a function of temperature for PbTe:K 1.25\%:Na 0.6\% (bullets), PbTe:K 1.25\%:Na 1\% (squares), PbTe:K 1.25\%:Na 1.4\% (triangles).  The high carrier density of the samples allows the calculation of the electronic part of $\kappa$ using a Lorenz number $L_{0}=(\pi^{2}/3)(k_{B}/e)^{2} \sim 2.45 \times 10^{-8} W \Omega Κ^{-2}$, through the Wiedemann-Franz law, $\kappa_{el}=L_{0} \sigma T$, for all temperatures of measurement.\cite{Ioffe} Subtracting the electronic part, $\kappa_{el}$, from $\kappa_{tot}$ we obtain the lattice part plotted with the corresponding open symbols on the same graph.  The $\kappa_{lat}$ of pure PbTe is also plotted (dotted line). At high temperatures, i.e. T$>$550K the extracted $\kappa_{lat}$ of the co-doped samples are not particularly low and follow reasonably well that of PbTe despite heavy doping with K and Na. The picture is quite different for T$<$500K, where it is evident that the extracted $\kappa_{lat}$ for PbTe:K 1.25\%:Na 1\% and PbTe:K 1.25\%:Na 1.4\% is higher than that for pure PbTe. The same holds true for other co-doped samples with Na content above 0.8\%. This indicates that the Lorenz number used in the calculation of $\kappa_{el}$ cannot correctly account for the electronic contribution to the thermal conductivity in this region. Instead, an $L$ value of $\sim$20-50\% higher than $L_{0}$ should be used at room temperature. This is another manifestation of interband scattering which evidently provides a rare mechanism of increasing considerably the Lorenz number above the fully degenerate value of $L_{0}$.\cite{Smirnov}  The higher $L$ number than $L_{0}$ implies the carriers transfer larger amounts of heat in the material than what is predicted by $L_{0}$. This involves the release of excess energy as heat by the carriers when they move from one band to the other. In fact, Kolomoets\cite{Kolomoets2} has shown that such an effect should be present whenever two bands coexist at the Fermi energy, like in semimetals, not necessarily restricted to bands of the same carrier type.\cite{Kolomoets2} 
 
\begin{figure}[hb]
\centerline{
\includegraphics[width=0.50\textwidth]{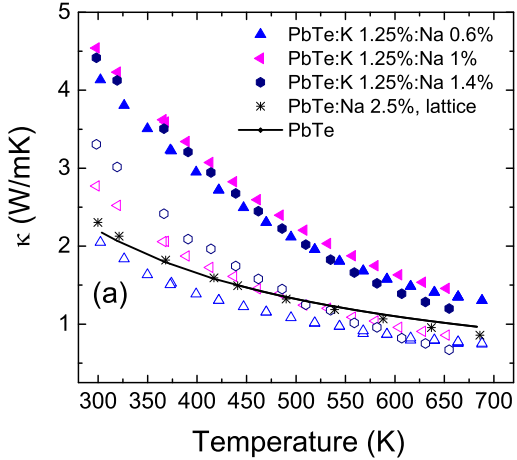}
}
\vspace{0.5cm}
\caption{\label{fig:Figure 7} Thermal conductivity, $\kappa$, as function of temperature for three co-doped samples: PbTe:K 1.25\%:Na 0.6\%, PbTe:K 1.25\%:Na 1\%, PbTe:K 1.25\%:Na 1.4\%. The filled symbols correspond to the total $\kappa$ and the corresponding open symbols to the extracted lattice part of $\kappa$. The black solid line corresponds to the lattice part of $\kappa$ of pure PbTe. The star symbols represent the extracted lattice thermal conductivity of a PbTe:Na sample with a hole density of $10 \times 10^{19} cm^{-3}$ at room temperature. Notice that for co-doped samples with increased Na content, in the range $300 \leq T \leq 500$ K, the lattice thermal conductivity deviates considerably from that of pure PbTe.
}
\end{figure}

\begin{figure}[hb]
\centerline{
\includegraphics[width=0.50\textwidth]{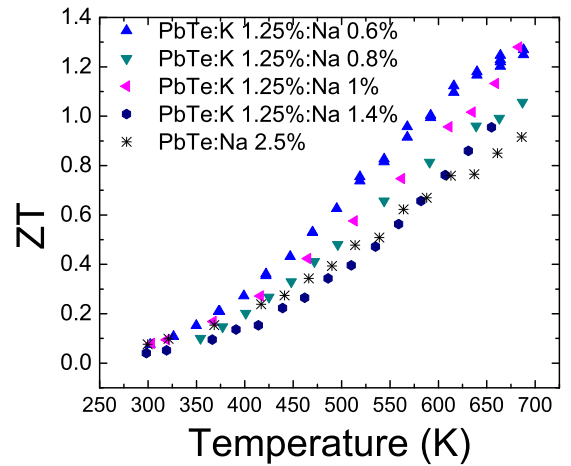}
}
\vspace{0.5cm}
\caption{\label{fig:Figure 8} Thermoelectric figure of merit, ZT, for selected K-Na co-doped samples, in comparison to the best PbTe sample doped with Na only. 
}
\end{figure}

Smirnov et al\cite{Smirnov} considered the case where carriers from band 1 transfer to band 2 giving up an energy $\Delta \varepsilon$. The resulting excess thermal conductivity from the two band interaction can be quantified by the relation:
\begin{eqnarray}
\kappa _{1-2}=\frac{\mathit{T\sigma }_{1}\sigma _{2}}{\sigma
_{1}+\sigma _{2}}(S_{2}-S_{1})^{2}
\label{eq:6}
\end{eqnarray}
                                                                                                                                                                  
Therefore, data in Fig. 7 include (a) the lattice contribution, (b) the band interaction contribution of Eq. \ref{eq:6}, (c) the carrier contribution from band 2, which is parabolic and should exhibit a Lorenz number $L_{0}$ and (d) the carrier contribution from band 1 which is non parabolic and the Lorenz number is a complex function of generalized Fermi integrals of the band gap and the reduced chemical potential.\cite{Smirnov} Clearly, the above considerations complicate the analysis for the extraction of a Lorenz number as a function of T, which is beyond the scope of this paper. The data in Fig 7(a) and 7(b) further support the argument that the interband scattering mechanism is at play for a wider temperature range in the co-doped samples as a result of both lattice expansion and doping in band 2. In contrast, crossing below the baseline in Fig. 7(b) around 425 K for the PbTe:Na ($\sim 10 \times 10^{19} cm^{-3}$) specimen supports previous suggestions that band 2 dominates transport in PbTe above $\sim$425K.\cite{Ravich-Chalcogenides} 

The resulting figures of merit for several of the codoped samples in this study are shown in Fig. 8. The maximum value measured reaches 1.3 at 700 K and this enhancement over conventional Na doped PbTe (ZT$\sim$0.9) is primarily associated with the enhanced power factors discussed above. 

\section{\label{sec:conc} Conclusions}

The Seebeck and Hall effect data of K, Na and K-Na doped PbTe do not support the development of resonant levels in the electronic band structure of PbTe but, rather, point to the importance of the heavy-hole valence band. Nevertheless, an enhancement in the thermoelectric properties similar in nature to what might be expected from resonance levels is observed and is attributed to the behavior of the heavy-hole valence band which responds to the lattice expansion with K doping and can be controlled to a significant degree by adjusting the Fermi energy with Na doping. Effectively, this causes the heavy-hole valence band to rise in energy and and adjust the Fermi energy and create a significant enhancement in the DOS that gives rise to interband scattering. This mimics the influence of a resonance level and has positive effects on the thermoelectric power factor.
 
As a result, ZT $\sim$ 1.3 is reached around 700K purely on account of the enhancement in the electronic properties and without a marked reduction in thermal conductivity.

\section{\label{sec:Ackn} Acknowledgements}

This work was supported by the Office of Naval Research (grant N00014-08-1-0613). The work at the University of Michigan is supported as part of the Revolutionary Materials for Solid State Energy Conversion, an Energy Frontier Research Center funded by the U. S. Department of Energy, Office of Basic Energy Sciences under Award Number DE-SC0001054. The work at Argonne National Laboratory is supported by Department of Energy, Office of Basic Energy Sciences (DE-AC02-06CH11357).

\end{document}